\documentclass[aps,prl,twocolumn,groupedaddress,showpacs,lettersize]{revtex4}
\usepackage{graphicx}

\begin{document}

\title{Vortex Formation by Interference of Multiple Trapped Bose-Einstein
Condensates}

\author{David R. Scherer, Chad N. Weiler, Tyler W. Neely, Brian P. Anderson}

\affiliation{College of Optical Sciences, University of Arizona,
Tucson, AZ 85721}

\date{October 5, 2006}

\begin{abstract}
We report observations of vortex formation as a result of merging
together multiple $^{87}$Rb Bose-Einstein condensates (BECs) in a
confining potential.  In this experiment, a trapping potential is
partitioned into three sections by a barrier, enabling the
simultaneous formation of three independent, uncorrelated
condensates. The three condensates then merge together into one BEC,
either by removal of the barrier, or during the final stages of
evaporative cooling if the barrier energy is low enough; both
processes can naturally produce vortices within the trapped BEC.  We
interpret the vortex formation mechanism as originating in
interference between the initially independent condensates, with
indeterminate relative phases between the three initial condensates
and the condensate merging rate playing critical roles in the
probability of observing vortices in the final, single BEC.

\end{abstract}


\maketitle

In a superfluid, long-range quantum phase coherence regulates the
dynamics of quantized vortices \cite{tilley1986sas,donnelly1991qvh}
and provides routes to vortex formation that are inaccessible with
classical fluids. For example, in dilute-gas Bose-Einstein
condensates (BECs), quantized vortices can be created using quantum
phase manipulation \cite{matthews1999vbe,leanhardt2002ivb}. Vortices
in BECs have also been created using methods more analogous to those
in classical fluid dynamics \cite{batchelor1980ifm}, namely through
rotating traps \cite{madison2000vfs,
aboshaeer2001ovl,hodby2001vnb,haljan2001dbe}, turbulence
\cite{inouye2001ovp}, and dynamical instabilities
\cite{anderson2001wds,dutton2001oqs}.  Yet in contrast with
classical fluid dynamics, to our knowledge vortex generation via the
mixing of initially isolated superfluids remains an unexplored
research area. Due to the availability and relative ease of
microscopic manipulation and detection techniques, BECs are
well-suited to address open questions regarding superfluid mixing
and associated vortex generation, along with the possible
accompanying roles of phase-coherence and matter-wave interference.

In this Letter, we describe experiments demonstrating that the
mixing or merging together of multiple condensates in a trap can
indeed lead to the formation of potentially long-lived quantized
vortices in the resulting BEC. We ascribe the vortex generation
mechanism to matter-wave interference between the initially
spatially isolated \emph{but otherwise identical} BECs, and show
that vortex formation may be induced even for slow mixing time
scales. While it is now well-known that matter-wave interference may
occur between BECs \cite{andrews1997oib}, our experiment
demonstrates a physical link between interference and vortex
generation, providing a new paradigm for vortex formation in
superfluids.  We emphasize that no stirring or phase engineering
steps are involved in our work, nor are any other means for
controllably nucleating vortices in the trapped atomic gas; the
vortex formation process itself is stochastic and uncontrollable,
and depends on relative condensate phases that are indeterminate
prior to condensate mixing.  The vortex formation mechanism
identified here may be particularly relevant when defects or
roughness are present in a trapping potential, or when multiple
condensates are otherwise joined together. Our experiment may also
illuminate aspects of vortex formation at site defects in other
superfluids, for which microscopic studies may be exceedingly
difficult and questions regarding vortex formation mechanisms are
unresolved.

\begin{figure}
\includegraphics{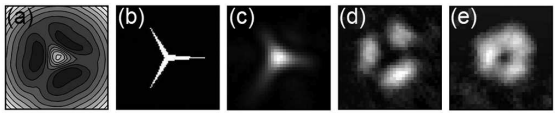}
\caption{\label{fig1} (a)  An illustration of the potential energy
profile of our three-well trap in a horizontal slice through the
trap center. This contour plot represents the addition of our
harmonic trap with the measured intensity profile of our optical
barrier beam, scaled to a potential energy. (b) A computer-generated
profile of the binary mask used to create our optical barrier, where
white represents the transmitting area. (c) An image of our optical
barrier. (d),(e) Phase-contrast images of trapped condensates
representing integrated optical depths along the vertical trap axis.
Each shows an area of 85 $\mu$m per side, as do (a) and (c). In (d),
three independent condensates are generated in the presence of a
strong barrier beam with 170 $\mu$W. (e) With 45 $\mu$W in the
barrier beam, the initial condensates have merged together during
evaporative cooling. Note that even this weak barrier displaces
atoms from the center of the BEC, so the observable density dip in
this image does not signify the presence of a vortex core.}
\end{figure}

To illustrate the basic concept underlying our experiment, we first
consider our atom trap, which is formed by the addition of a
time-averaged orbiting potential (TOP) trap \cite{petrich1995stc}
and a central repulsive barrier created with blue-detuned laser
light that is shaped to segment the harmonic oscillator potential
well into three local potential minima.  Figure \ref{fig1}(a) shows
an example of potential energy contours in a horizontal slice
through the center of our trap. We will assume throughout the
ensuing descriptions that the energy of the central barrier is low
enough that it has negligible effect on the thermal atom cloud; such
is the case in our experiment. However, the central barrier does
provide enough potential energy for an independent condensate to
begin forming in each of the three local potential minima from the
one thermal cloud.  If the central barrier is weak enough,
condensates with repulsive interatomic interactions will grow
together during evaporative cooling; if the barrier is strong
enough, the condensates will remain independent. In this latter
case, the central barrier height may be lowered while keeping the
condensed atoms held in the TOP trap.  Overlap and interference
between the heretofore independent condensates would then be
established as the condensates merge together into one.  We have
examined both scenarios.

Depending on the relative phases of the three interacting
condensates and the rate at which the condensates merge together
(via either process), the final merged BEC may have nonzero net
angular momentum about the vertical trap axis, as we now describe.
We first recall that the relative phase between two independent
superfluids is indeterminate until an interference measurement is
made. However, when interference occurs, a directional mass current
will be established between the superfluids. A relative phase can
then be determined, but it will vary randomly upon repeated
realizations of the experiment
\cite{anderson1986,javanainen1996qpb}. In our experiment, when the
initial condensates merge together, fluid flow in the intervening
overlap regions is established; a straightforward model of the mass
current for two overlapping but otherwise uncorrelated states shows
that the direction of fluid flow depends on the sine of the relative
phase between the states \cite{SUPERPOS}.

When our three condensates are gradually merged together while
remaining trapped, fluid flow that is simultaneously either
clockwise or counter-clockwise across all three barrier arms may
occur with finite probability. For ease of this discussion, and
keeping in mind that only relative phases carry physical meaning, we
imagine that the condensates formed in the three local minima of
Fig.\ \ref{fig1}(a) can be labeled with phases $\phi_j,$ where the
indices $j$=1, 2, and 3 identify the condensates in a clockwise
order, respectively. Upon merging of the three condensates, if it
turns out that for example, $\phi_2 - \phi_1 = 0.7\pi$ and $\phi_3 -
\phi_2 = 0.8\pi$ (thus necessarily $\phi_1 - \phi_3 = 0.5\pi$ since
each $\phi_j$ must be single valued), then some finite amount of
clockwise-directed fluid flow will be established between each pair,
hence also for the entire fluid. More generally, if the three
merging condensates have relative phases $\phi_2 - \phi_1$, $\phi_3
- \phi_2$, and $\phi_1 - \phi_3$ that are each simultaneously
between 0 and $\pi$, or each between $\pi$ and 2$\pi$, the resulting
BEC will have nonzero angular momentum, which will be manifest as a
vortex within the BEC. By examining the full range of phase
difference possibilities, the total probability $P_v$ for a net
fluid flow to be established in either azimuthal direction is
determined to be $P_v=0.25$, given statistically random phase
differences for each experimental run. $P_v$ is thus the probability
for a vortex to form as the three condensates merge together.

Absent from the above description is an analysis of the phase
\emph{gradients} of the three condensates during the merging
process, which will lead to transient interference fringes in
overlapping condensates. These fringes may decay to numerous
vortices and antivortices, and possibly vortex rings, similar to
instabilities of dark solitons in BECs
\cite{anderson2001wds,dutton2001oqs,feder2000dss,THRY}.  For rapidly
merged condensates, we may thus expect the observation of multiple
vortex cores in an BEC image, or to find a value of $P_v$ greater
than 0.25.  Yet as the condensates are merged together more and more
slowly, the dependence of $P_v$ on phase gradients becomes
negligible and $P_v$ should approach a limiting value of 0.25.

Our experiment is designed to study the presence of vortices in an
$|F$=1, $m_F$=$-1\rangle$ $^{87}$Rb BEC subsequent to the merging of
three condensates such created in a three-segment potential well. To
create just a single BEC in a trap without a central barrier, we
first cool a thermal gas to just above the BEC critical temperature
in an axially symmetric TOP trap with radial (horizontal) and axial
(vertical) trapping frequencies of 40 Hz and 110 Hz, respectively.
We then ramp the TOP trap magnetic fields such that the final trap
oscillation frequencies are $7.4$ Hz (radially) and $14.1$ Hz
(axially). A final 10-second stage of radio-frequency (RF)
evaporative cooling produces condensates of $\sim$4$\times$10$^5$
atoms, with condensate fractions near 65\% and thermal cloud
temperatures of $\sim$22 nK. The chemical potential of such a BEC is
$k_B \times$8 nK, where $k_B$ is Boltzmann's constant.

To create instead three isolated condensates in a segmented trap, we
modify the above procedure by ramping on an optical barrier
immediately before the final 10-s stage of evaporative cooling in
the weak TOP trap. The barrier itself is formed by illuminating a
binary mask, illustrated in Fig.\ \ref{fig1}(b), with a focused
blue-detuned Gaussian laser beam of wavelength 660 nm. After passing
through the mask, the beam enters our vacuum chamber along the
vertical trap axis.  The mask is imaged onto the center of the atom
cloud with a single lens. Due to diffraction, the beam has an
intensity profile as shown in Fig.\ \ref{fig1}(c), with a maximum
intensity and thus barrier energy aligned with the center of the TOP
trap. The barrier's potential energy decreases to zero over $\sim$35
$\mu$m along three arms separated by azimuthal angles of
approximately $120^{\circ}$. With 170 $\mu$W in the beam, which
corresponds to a maximum barrier height of $k_B \times$26 nK for our
beam, three condensates are created and do not merge together during
their growth; a set of three BECs created under these conditions is
shown in Fig.\ \ref{fig1}(d). With instead 45 $\mu$W in the beam,
corresponding to a maximum barrier energy of $k_B \times$7 nK, three
independent condensates again \emph{initially} form, but as the
condensates grow in atom number, their repulsive interatomic
interactions eventually provide enough energy for the three
condensates to flow over the barrier arms.  The three condensates
thus naturally merge together into one BEC during evaporative
cooling, as shown in Fig.\ \ref{fig1}(e).

In our first study, three spatially isolated condensates were
created in the presence of a strong barrier of maximum potential
energy $k_B \times$26 nK, and were then merged together by ramping
down the strength of the barrier to zero linearly over a time
$\tau$.  Any vortex cores formed by this process in the resulting
BEC have a size below our optical resolution limit, and are too
small to be directly observed in the trapped BEC.   We thus suddenly
removed the trapping potential and observed the atom cloud using
absorption imaging along the vertical axis after 56 ms of ballistic
expansion.  This entire process was repeated between 5 and 11 times
for each of 6 different values of the barrier ramp-down time $\tau$
between 50 ms and 5 s.

\begin{figure}
\includegraphics{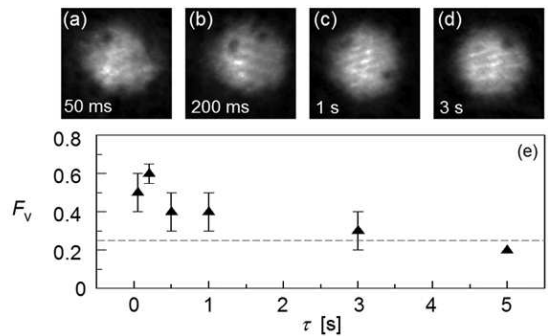}
\caption{\label{fig2} (a)--(d) Images showing vortices in
condensates created in the presence of a strong ($k_B \times$26 nK)
barrier. Prior to release from the trap, the barrier was ramped off
over the time $\tau$ indicated. Each absorption image shows a region
of 170 $\mu$m per side. (e) Vortex observation fraction $F_{v}$ vs.\
barrier ramp-down time $\tau$. The data for $\tau$ values of 50 ms,
100 ms, 200 ms, 500 ms, 1 s, 3 s, and 5 s, consisted of 5, 11, 10,
10, 10, 5, and 5 images, respectively.  For clarity, statistical
uncertainties due to finite sample sizes are not reflected in this
plot, but they generally exceed our counting uncertainties.}
\end{figure}

In a significant fraction of our experimental runs, we observed one
or more vortex cores in our condensates, a clear indication that
condensate merging can indeed induce vortex formation. Moreover, the
observed spatial density distributions vary from shot to shot, as
would be expected with indeterminate phase differences between the
initial condensates.  However, many images are absent of any
vortices.  Example images of expanded BECs in Fig.\
\ref{fig2}(a)--(d) show the presence of vortex cores after various
barrier ramp-down times. An analysis of vortex observation
statistics is given in Fig.\ \ref{fig2}(e) for the different values
of $\tau$ examined. We define a vortex observation fraction $F_{v}$
as the fraction of images, for each value of $\tau$, that show at
least one vortex core \cite{RINGS}. The error bars reflect the
ambiguity in our ability to determine whether or not an image shows
at least one vortex. For example, core-like features at the edge of
the BEC image, or core-like features obscured by imaging noise, may
lead to uncertainty in our counting statistics and determination of
$F_{v}$.  As the plot shows, $F_{v}$ reaches a maximum near 0.6 for
the smaller $\tau$ values, and drops to $\sim$0.25 for long
ramp-down times.  We expect that with larger sample sizes, $F_{v}$
should approximate $P_v$ for each $\tau$.  Thus our results are
consistent with our conceptual analysis, where $P_v > 0.25$ for fast
merging times when interference fringes may occur, and $P_v=0.25$
for long merging times.

For $\tau \leq 1$ s, multiple vortices were often observed in our
images, as images of Fig.\ \ref{fig2} show, indicating that phase
gradients are likely to play an important role in vortex formation
if condensates are quickly merged together. Furthermore,
observations of multiple vortex cores may indicate the presence of
vortices and antivortices. Although we are unable to determine the
direction of fluid circulation around our observed vortex cores, we
performed a test in which the barrier was ramped off in 200 ms, thus
forming multiple vortex cores with a high probability.  We then
inserted additional time to hold the final BEC in the unperturbed
harmonic trap before our expansion imaging step. After such a
sequence, the probability of observing multiple vortices dropped
dramatically: for no extra hold time, we observed an average of 2.1
vortex cores per image, whereas this number dropped to 0.7 for an
extra 100 ms hold time, suggestive of either vortex-antivortex
combination or other dynamical processes by which vortices leave the
BEC. However, images showing single vortices were observed even
after 5 s of extra hold time in our trap following the 200 ms
barrier ramp, indicating a relatively long vortex lifetime in our
trap.

\begin{figure}
\includegraphics{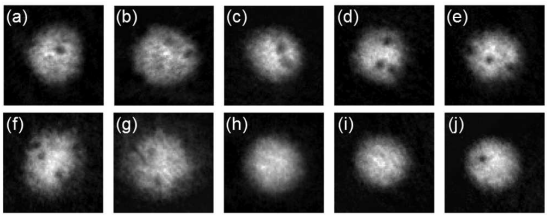}
\caption{\label{fig3} (a),(b) Images showing the presence of
vortices naturally occurring in condensates created in a potential
well with a $k_B \times$7 nK segmenting barrier. (c)--(h) Images
obtained using intermediate barrier heights. (i),(j) Images taken in
the absence of a central optical barrier. Note the presence of a
spontaneously formed vortex in (j). All absorption images represent
a region of 170 $\mu$m per side.}
\end{figure}

In a second investigation, we differed from the above experiment by
using a weaker barrier with a maximum energy of $k_B \times$7 nK
such that the three condensates naturally merged together into one
BEC during the evaporative cooling process.  We emphasize that this
merging process is due solely to the increasing chemical potentials
exceeding the potential energy of the barrier arms between the
condensates; the barrier strength remained constant throughout the
growth and merging of the condensates when vortices may form.  After
evaporative cooling, we ramped off the optical barrier over 100 ms
and released the atoms from the trap to observe the BEC after
ballistic expansion. Under these conditions, our vortex observation
fraction was $F_{v}$=0.56$\pm$0.06 in a set of 16 images, with
example images shown in Fig.\ \ref{fig3}(a) and (b).

By adding an extra 500 ms of hold time after the final stage of BEC
formation but \emph{before} the start of the 100 ms barrier
ramp-down and ballistic expansion, the vortex observation fraction
decreased to $F_{v}$=0.28$\pm$0.14. Again, this drop in probability
may be due to vortex-antivortex combination during extra hold time
in the weakly perturbed harmonic trap. We thus conclude that with a
maximum barrier energy of $k_B \times$7 nK, vortices are formed
during the BEC creation process rather than during the ramp down of
the weak barrier, consistent with our phase-contrast images of
trapped BECs that show a ring-like rather than segmented final
density distribution.

Barrier strengths between the two limits so far described also lead
to vortex formation, either during BEC growth or during barrier ramp
down.  With a barrier strength in this range, up to at least four
clearly defined vortex cores have been observed in single images, as
the examples of Fig.\ \ref{fig3}(c)--(f) show. Density defects other
than clear vortex cores have also appeared in our images, as in the
upper left of Fig.\ \ref{fig3}(g), where a ``gash''-like feature may
be a possible indicator of vortex-antivortex combination; similar
features have been seen in related numerical simulations
\cite{THRY}. Often, however, no vortices are observed; an example
with no vortex cores appearing is Fig.\ \ref{fig3}(h). For
comparison, an expansion image taken after creating a condensate in
a trap without a barrier is shown in Fig.\ \ref{fig3}(i).

Perhaps surprisingly, single vortex cores have also appeared in
$\sim$10\% of our expansion images taken in the absence of any
central barrier. In other words, for our basic single BEC creation
procedure as outlined above, and \emph{without a segmenting barrier
ever turned on}, vortices occasionally form spontaneously and are
observable in expansion images.  An example image is shown in Fig.\
\ref{fig3}(j). These observations may be related to predictions of
spontaneous vortex formation due to cooling a gas through the BEC
transition \cite{zurek1985ces}. We are currently investigating these
intriguing observations further.

We finally note that to generate vortices by the mechanism described
in this paper, it is important for two reasons that condensates
merge and interfere while trapped, as opposed to during expansion.
First, in a trapped BEC, the nonlinear dynamics due to interatomic
interactions would play a key role in the structural decay of
interference fringes, which may be responsible for generation of
multiple vortices and antivortices seen with fast merging times.  In
an expanding gas, the interactions become negligible as the gas
density decreases. Second, by keeping condensates trapped during
their mixing, we are able to study slow merging, where we believe
that relative overall phases are primarily responsible for vortex
generation.  We conjecture that with slow merging, it may be
possible to directly imprint relative phases onto three or more
trapped, separated, and phase-correlated condensates to controllably
engineer vortex states.

In summary, we have generated vortices by merging together isolated
and initially uncorrelated condensates into one final BEC. We have
shown that our vortex observation statistics are consistent with a
simple conceptual theory regarding indeterminate phase differences
between the initial condensates; however, quantitative theoretical
examination is needed for further analysis of our results. We have
also demonstrated that condensates created in the presence of weak
trapping potential defects or perturbations, such as our weak
optical barrier, may naturally acquire vorticity and nonzero orbital
angular momentum. This result challenges the common notion that a
BEC \emph{necessarily} forms with no angular momentum in the lowest
energy state of a trapping potential;  rather, the shape of a static
confining potential may be sufficient to induce vortex formation
during BEC growth, a concept that may be of relevance to other
superfluid systems as well.

We thank Ewan Wright and Poul Jessen for helpful discussions, and
Tom Milster for use of his Maskless Lithography Tool to create our
optical barrier mask and phase plates for phase-contrast imaging.
This work was funded by grants from the ARO and NSF.



\begin{thebibliography}{26}
\bibitem[{SUP()}]{SUPERPOS}
\bibinfo{note}{For example, in the superposition $\Psi(x) =
A\cdot \{\exp[-(x-x_0)^2 + i\phi_1] + \exp[-(x+x_0)^2 + i\phi_2]\}$,
where $A$ is a normalization coefficient and $x$ is a dimensionless
spatial variable, the mass current density at $x=0$ is proportional
to $\sin{(\phi_1 - \phi_2)}$.}
\bibitem[{THR()}]{THRY}
\bibinfo{note}{We have numerically simulated three condensates merging together
  in a quasi-two-dimensional geometry, finding results in qualitative agreement
  with our experiment and our conceptual arguments. These numerical results
  will be described in a forthcoming publication.}
\bibitem[{RIN()}]{RINGS}
\bibinfo{note}{Since we are unable to clearly distinguish between vortex rings and vortex lines,
it is possible that some features that we count as vortices are
actually vortex rings. }
\end{thebibliography}
\end{document}